 \documentclass[10pt,twocolumn,aps,showpacs]{revtex4}

 \usepackage{graphicx}

\begin{document}


 \title{Two Particle Azimuthal Correlations in 4.2A GeV C+Ta Collisions}


 \author{Lj. Simi\'c $^1$}
 \email[]{simic@phy.bg.ac.yu}
 \author{O. Jotanovi\'c $^2$}
 \author{J. Milo\v sevi\'c $^3$}
 \author{I. Menda\v s $^1$}

\affiliation{$^1$ Institute of Physics, P.O. Box 68, 11081 Belgrade, Serbia\\
$^2$ Faculty of natural science, Banja Luka, Bosnia and Herzegovina\\
 $^3$ Department of physics, University of Oslo}


\date{\today}

 \begin{abstract}
 Two particle azimuthal correlations are studied
 in 4.2A GeV C+Ta collisions observed with the 2-m propane bubble chamber
 exposed at JINR Dubna Synchrophasotron.
 The correlations are analyzed both  for protons
 and negative pions, and their dependence on the collision centrality,
 rapidity and rapidity difference is investigated.
   It is found that  protons show
  a weak back-to-back correlations, while a side-by-side
 correlations are observed for negative pions.
   Restricting both protons to the target or projectile fragmentation region,
 the side-by-side correlations are observed for protons also.
Using the two particle
 correlation function, the flow analysis is performed and intensity of
 directed flow is determined without event-by event estimation of the reaction plane.

 \end{abstract}


\maketitle


  The measurement of azimuthal ($\Delta \phi$) correlations for
  charged hadrons is arguably one of the most important probes for
  high density nuclear matter created in relativistic nucleus-nucleus
  collisions ~\cite{Liu}. Detailed and systematic measurements constitute an 
  important
  step in good understanding of their mechanistic origin and  provide
  additional information to the methods, currently used to measure azimuthal
  anisotropies, which  assume that all azimuthal correlations
  between particles result from their correlation with the reaction
  plane ~\cite{Flow1,Flow2,Flow3,Flow4,Flow5}.
     
  Here, we present a  study of two particle azimuthal correlations
  in 4.2A GeV C+Ta collisions. The results are obtained by analyzing 2000
  C+Ta events recorded with the 2-m propane bubble chamber exposed at JINR
  Dubna Synchrophasotron. Additionally, the same type of analysis  is
  performed using 50000 events generated by the Quark Gluon String
  (QGS) model ~\cite{QGS}. The two particle azimuthal correlations are
  studied both  for protons and negative pions, and their dependence
  on the collision centrality, rapidity and rapidity difference
  is investigated. By assuming that azimuthal correlations between two
  particles are generated by the correlation of the azimuth of each particle
  with the reaction plane, a directed flow of  negative pions and protons
  is determined from two particle azimuthal correlations.
    
    To study the collisions with tantalum nucleus $^{181}$ Ta, three tantalum
  foils (1 mm thick and 93 mm  apart) were placed inside the chamber
  working in the 1.5 T magnetic field. The characteristics of the
  chamber allow precise determination of the multiplicity and momentum of all
  charged particles, as well as identification of all negative and
  positive particles with momenta less than 0.5 GeV/c. All  recorded negative
  particles, except identified electrons, are taken to be $\pi^-$. Among
  them remains admixture of unidentified fast electrons ($<$ 5\%).
  All positive particles with momenta less than 0.5 GeV/c
  are classified either as protons or $\pi^+$ mesons according to their
  ionization density and range. Positive particles with momenta above
  0.5 GeV/c are taken to be protons, and because of this, the admixture
  of $\pi^+$ of approximately 7\%  is subtracted statistically using
  the procedure described in ~\cite{Simic}. From the resulting number
  of protons, the projectile spectators (protons with momenta
  $p>$ 3 GeV/c and emission angle $\theta<4^{\rm o}$) and target
  spectators (protons with momenta $p<$ 0.3 GeV/c) are further subtracted.
  The experimental data are also corrected to the loss of particles
  emitted at small angles relative to the optical axes of chamber and to the
  loss of particles absorbed by the tantalum plates. The aim of this
  correction is to obtain isotropic distribution in azimuthal angle and
  smooth distribution in emission angle (both measured with respect
  to the direction of the incoming projectile).
 
   Following the approach commonly used in interferometry studies, a
  two particle azimuthal correlation function 
   ~\cite{Wang,Lacey1, Lacey2, Borg1}
  is used  to measure the distribution of the azimuthal angle difference
  between pairs of charged hadrons:
      $$C(\Delta\phi)= {{N_{cor}(\Delta\phi)}\over {N_{mix}(\Delta\phi)}},$$
  where $\Delta \phi$ represents the angle
  between the transverse momenta of {\it i}-th and {\it j}-th
  particle in an event:
   $$ \Delta \phi = \arccos \frac{{\bf p}_{Ti} \cdot {\bf p}_{Tj}}
      { p_{Ti}\ p_{Tj}}. $$
  Here $N_{cor}(\Delta\phi)$ represents the experimentally 
  observed $\Delta\phi$
  distribution for particle pairs selected from the same event, and
  $N_{mix}(\Delta\phi)$  represents the $\Delta\phi$ distribution for
  uncorrelated particle pairs selected from the mixed events. Mixed events
  were obtained by randomly selecting each  member of particle pair from
  different events having similar multiplicity.
  Essentially, $C(\Delta\phi)$  measures whether the particles  are
  preferentially emitted  back-to back ($C(\Delta\phi) >1$ at
  $\Delta\phi\approx180^0$) or side-by-side ($C(\Delta\phi) >1$ at
  $\Delta\phi\approx 0^0$)   in the transverse plane.
  \begin{figure}
  \includegraphics [width=6.3cm] {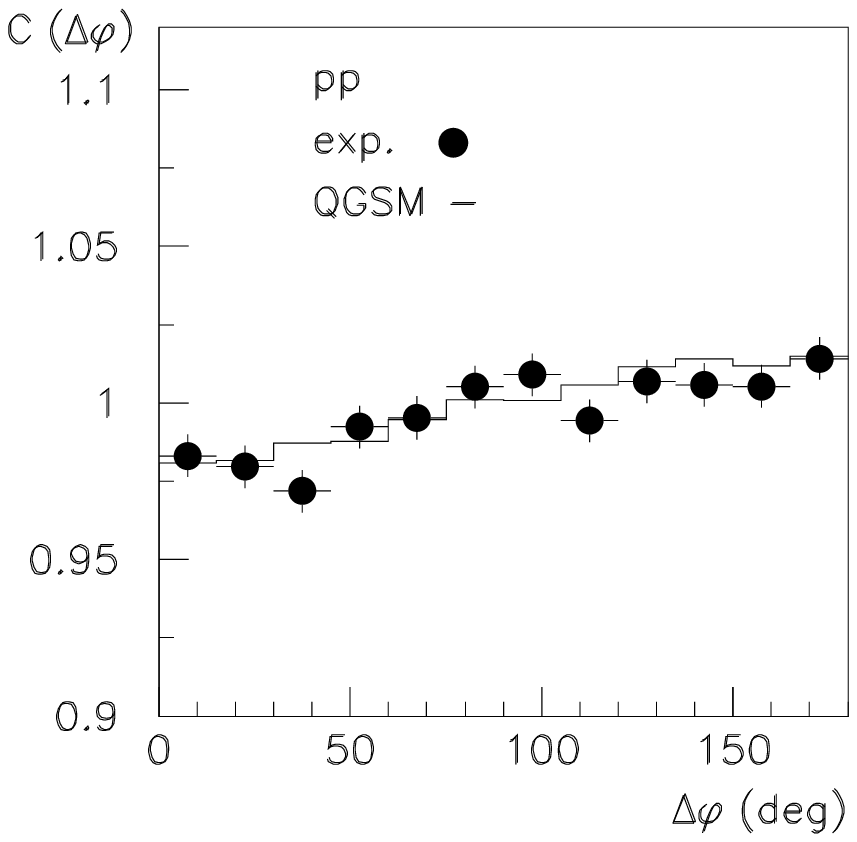}
  \includegraphics [width=6.3cm] {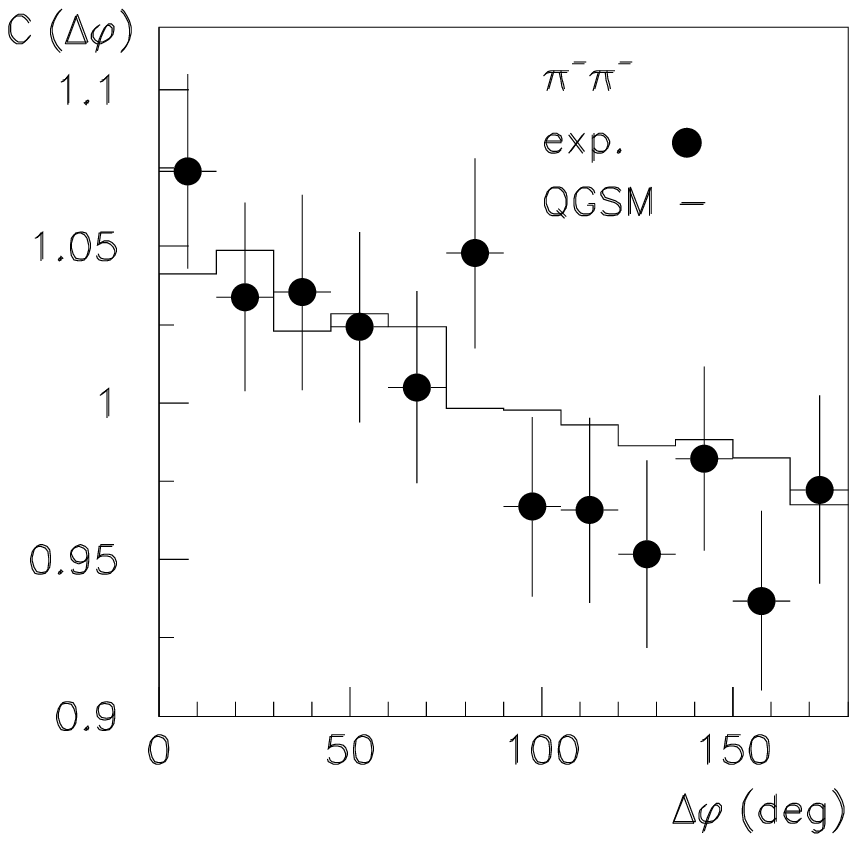}
  \caption{Two particle azimuthal correlations for
   $pp$ and $\pi^- \pi^-$  for experimental data ({\it full circles}) and QGSM ({\it histograms}).}
 \end{figure}
  
  Figure 1 shows the two particle azimuthal correlations
  for proton pairs  and negative pion pairs  for experimental
  data  and QGSM. For proton and pion pairs 
  clear correlations are observed,
  but the correlation curves have opposite slopes.
  Protons are produced predominantly back-to back, meaning that
  correlation function $C(\Delta\phi)$ has maximum at $\Delta \phi\sim 180^0$ and
  minimum at $\Delta \phi\sim 0^0$.
  These back-to-back correlations are stronger for peripheral collisions
  and decrease when going to semicentral collisions. Peripheral collisions are extracted
  by selecting $\approx 50\%$ of the total number of events with the lowest
  multiplicity of the number of participant protons ($N_p<11$).
  The QGSM reproduces slope and magnitude of correlations,
  as well as its dependence on collision centrality.
  Pions are produced  predominantly side-by-side with
  $C(\Delta\phi) >1$ at small $\Delta \phi$.
  The QGSM reproduces this structure of correlation function,
  but somewhat underestimates the magnitude of correlations.
  The QGSM also predicts decrease in magnitude of correlations
  when going from peripheral to semicentral collisions,  while 
  the statistics of the
  experimental data is too low to observe any clear dependence
  of pion correlations on the collision centrality.
 \begin{figure*} 
  \includegraphics [width=5.6cm] {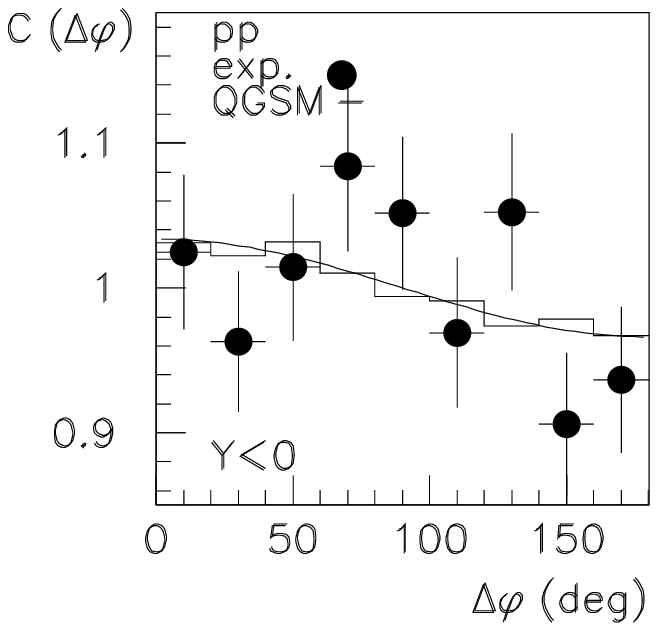}
 \includegraphics [width=5.6cm] {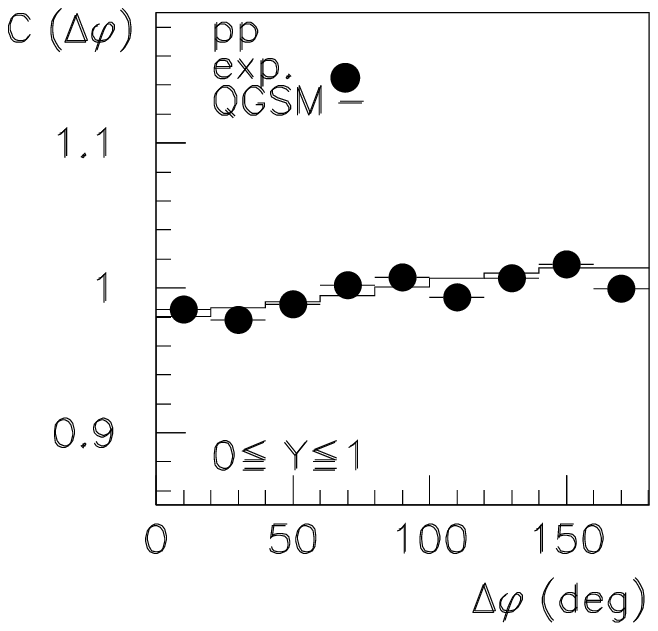}
 \includegraphics [width=5.6cm] {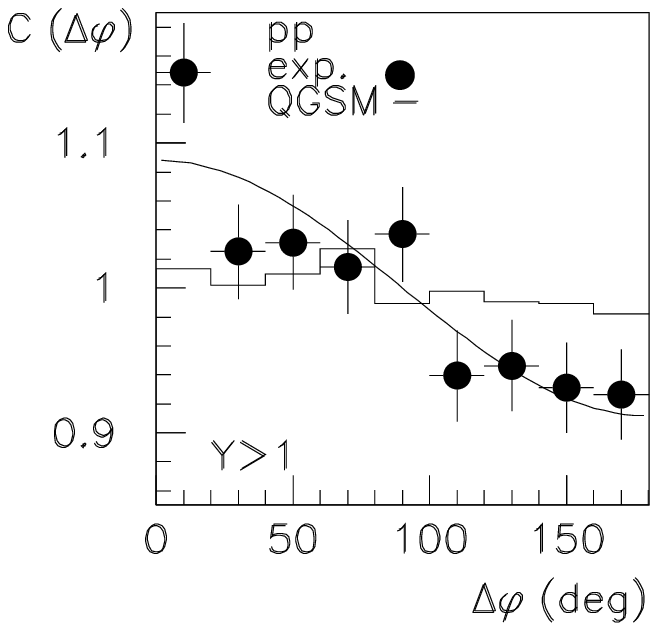}\\
 \includegraphics [width=5.6cm] {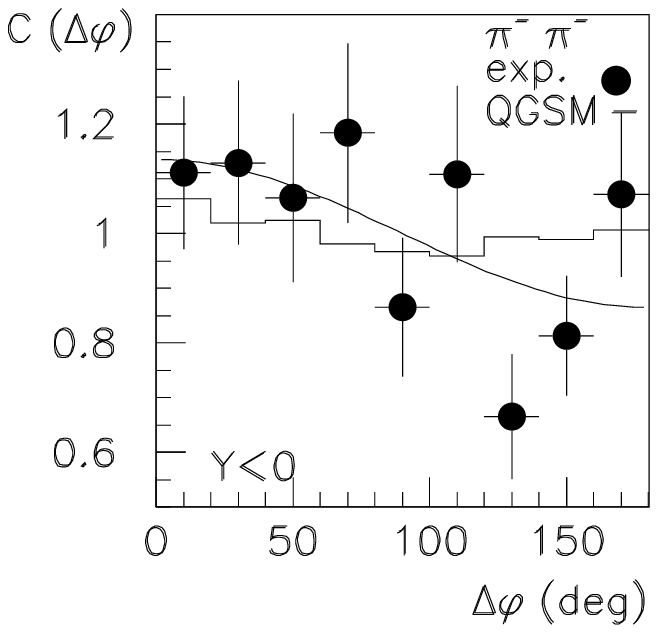}
 \includegraphics [width=5.6cm] {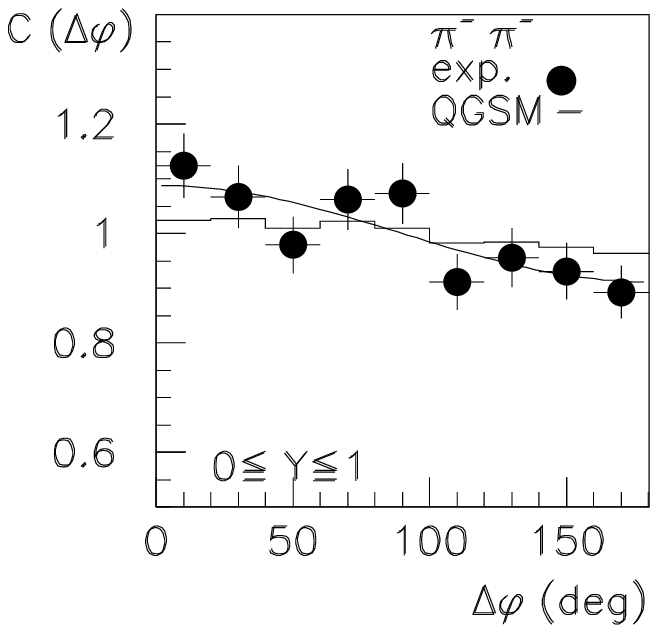}
 \includegraphics [width=5.6cm] {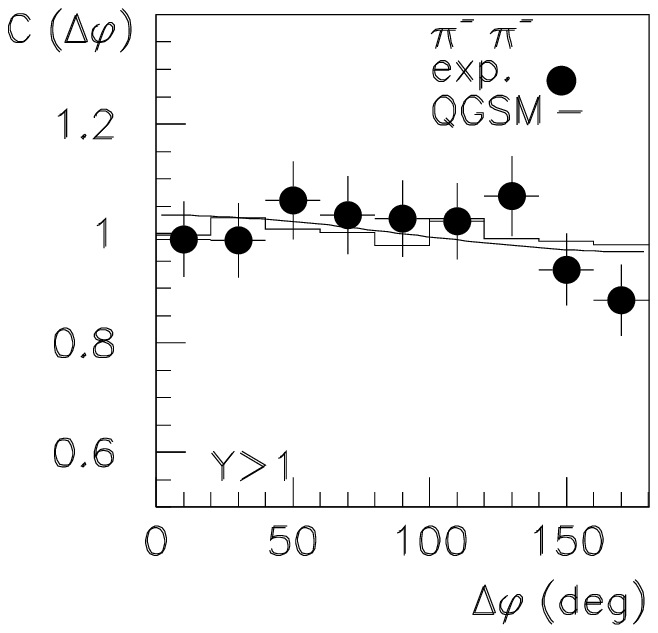}\\
  \caption{Correlation function for $pp$ and $\pi^-\pi^-$ pairs, 
  for three  rapidity     intervals: $y<0$, $0\leq y \leq 1$ and $y>1$.}
   \end{figure*}
  
  The azimuthal correlations for particles  emitted in the  selected
  rapidity intervals, $y<0$, $0\leq y \leq 1$
  and $y>1$, are shown in Fig. 2. These three intervals  roughly  correspond to
  the target fragmentation region, mid rapidity region and projectile
  fragmentation region, respectively. In the target and projectile rapidity region
  protons show side-by-side correlations, while in the mid rapidity region
  a very weak back-to-back  correlations still exist. The QGSM reproduces these
  experimental results, although  underestimates
  the magnitude  of proton correlations in the projectile rapidity region.
  For negative pions, side-by-side correlations exist in all three rapidity
  intervals. However, there is a slight depletion of the
  magnitude of correlations when going from the target to the projectile
  rapidity region. Similar side-by-side correlations
  were also observed for positive pions at 4.9 GeV p+Au, 200A GeV/c
  p+Au, O+Au and S+Au collisions ~\cite{Awes}.
  For charged pions, at
   1A GeV/c Au+Au collisions, an anisotropic pion flow relative
   to the reaction plane was found ("pion squeeze-out") ~\cite{Brill}.
   The side-by-side correlations in ~\cite{Awes}
  increase both with target mass and with impact parameter of a 
  collision and were
  consistently described by assuming strong rescattering phenomena 
  including pion absorption effects in the entire excited target nucleus.
   However, for 1A GeV/c Au+Au collisions, due to different reaction 
   geometry in symmetric systems and a longer nuclear passing time, 
   the spectator matter causes a stronger
   absorption and rescattering in reaction plane than out of 
   reaction plane causing an appearance of the out-of-plane pion flow.
 \begin{figure}
\includegraphics [width=2.8cm] {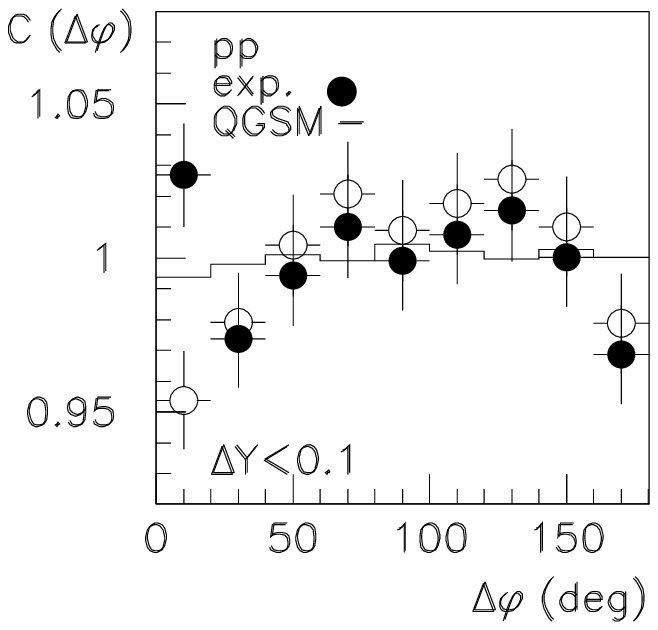}
\includegraphics [width=2.8cm] {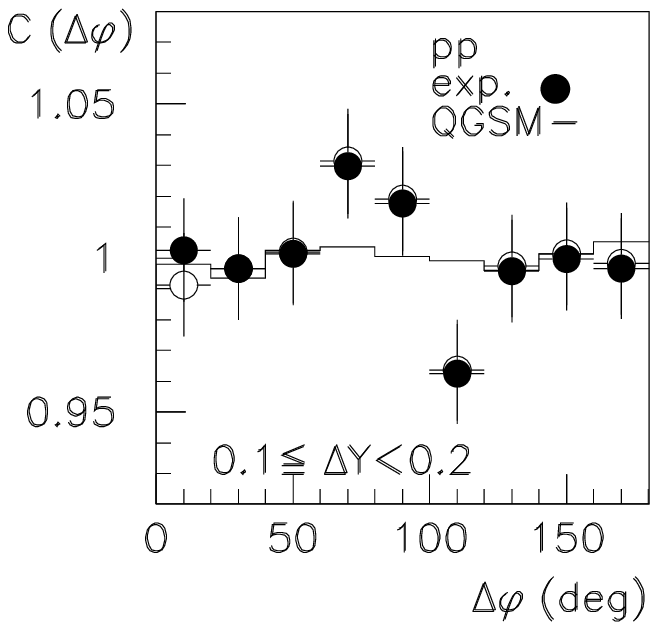}
\includegraphics [width=2.8cm] {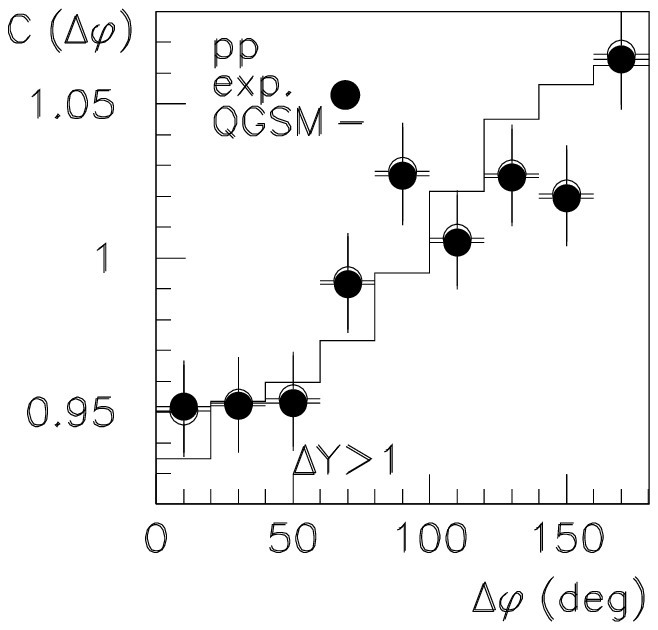}\\
\includegraphics [width=2.8cm] {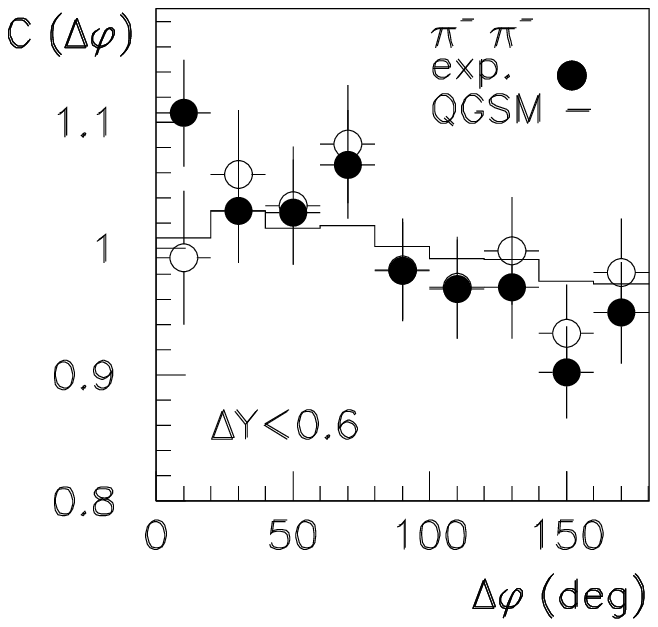}
\includegraphics [width=2.8cm] {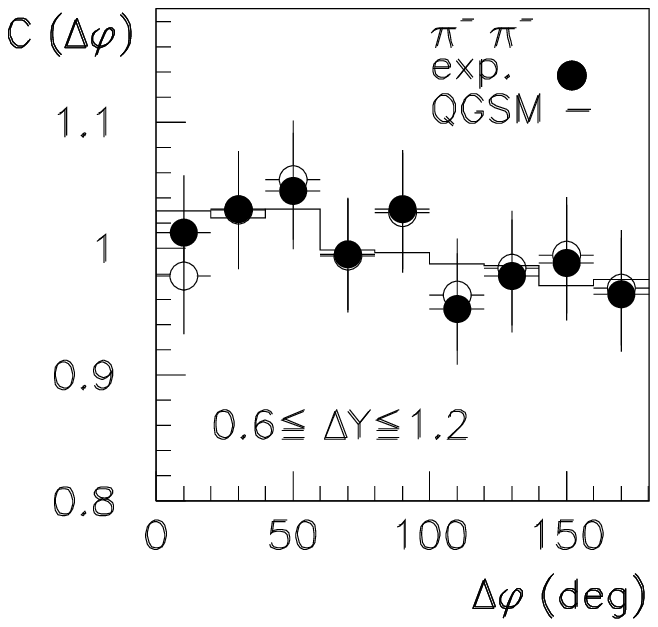}
\includegraphics [width=2.8cm] {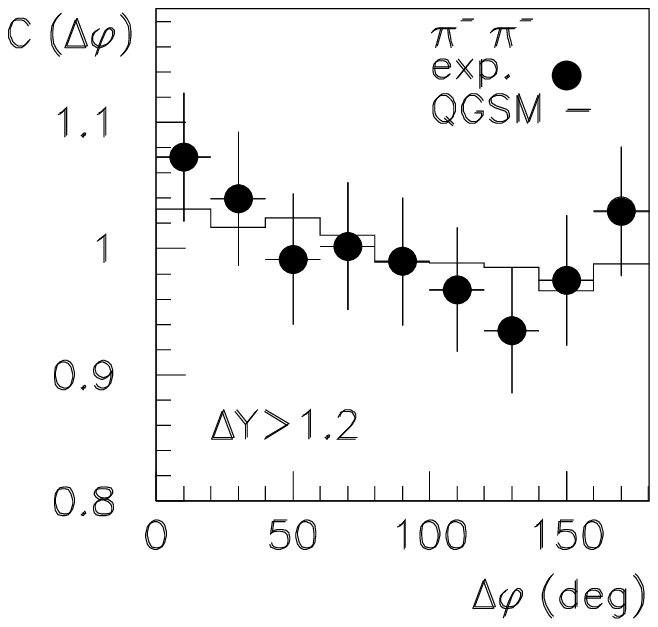}\\
  \caption{Correlation function for $pp$ and $\pi^-\pi^-$ pairs, for 
  various $\Delta y$  values with  ({\it full circles}) and without  
  ({\it open circles}) contributions from  quantum correlations.}
 \end{figure}
  The dependence of azimuthal correlations on the rapidity difference
  $\Delta y=y_1-y_2$, between   particles in a pair, is shown in Fig. 3.
  For  protons emitted with $\Delta y>1$, correlation function has a clear
  back-to-back  structure. With decreasing $\Delta y$  the slope of  $C(\Delta\phi)$
  decreases, and for protons emitted with  $\Delta y<0.2$ the correlation function
  is flat within error bars. For $\Delta y<0.1$, on this flat structure
  a superimposed  peak at $\Delta \phi<20^0$ appears. This peak
  can be attributed to the correlations arising from the effect of quantum
  (Fermi-Dirac) statistics, and final state interactions due to strong
  and Coulomb force. The peak disappears if protons with
  $q_T< 0.07$ GeV/c are removed ~\cite{HBT}, where ${\bf q=p_1- p_2}$
  and ${\bf q_T}$ is the component of {\bf q}  in the direction
  perpendicular to ${\bf p_1+p_2}$. The QGSM  reproduces
  changes of $C(\Delta\phi)$ with $\Delta y$ except the peak due to
  close pairing of protons.

 For  pions, the observed side-by-side correlations are roughly independent of
 the value of the rapidity difference.
 However, effect of close pairing  of $\pi^-$ mesons,
 due to Bose-Einstein statistics, can also be observed
 in the structure of correlation function at $\Delta \phi<20^0$  for
 $\Delta y<0.6$  (Fig. 3  bottom). This peak disappears  if pions with
 $q \leq 0.160$ GeV are removed ~\cite{HBT}.
 Since 11\% pion  and 0.8\% proton pairs contribute to quantum correlations,
 their effect on overall azimuthal correlations is negligible.

 Assuming that all  side-by-side  correlations between two particles
 are generated  by the correlations of the azimuth of each particle with
 the reaction plane, we can extract the magnitude of the directed flow $v_1$
 by fitting $C(\Delta\phi)$ with ~\cite{Lacey2, Borg1}:
 $$C(\Delta\phi)\propto 1+ 2v_1^2 \cos(\Delta\phi).$$
 The fit of the correlation function is shown in Fig.2, while
 the extracted values of the directed flow are summarized in Table 1.
 
 The extracted values of directed flow can be compared with the values
 ~\cite{Simic} obtained from the Fourier expansion of the azimuthal
 distribution of particles with respect to the reaction plane.
 The latter method predicts that for $y>1$ ($y<0$) the protons show
 a positive (negative) directed flow with magnitude $v_1\simeq 0.17$
 ($v_1\simeq -0.07$).
 For $0\leq y \leq 1$ the coefficient $v_1$ changes the sign, with the
 zero crossing at $y\simeq0.5$ that corresponds to the average rapidity
 of protons.
 We can see that both methods give consistent values of the directed
 flow, with exception that from  two particle azimuthal correlations we
 can not determine the sign of the $v_1$ coefficient.
 For the pions, the method
 which involves the reaction plane gives  positive $v_1$  for all rapidities.
 In the  region $0\leq y \leq 1$, where the statistics is largest,
 one has $v_1\simeq 0.10$. Comparing the results of both methods we see that
 two particle azimuthal correlations overestimate the flow of pions.
\begin {table}
\begin{center}
\caption{Magnitude of directed flow, $v_1$, of protons and negative pions 
determined from the two-particle
 azimuthal correlation functions.}
 \begin{tabular}{l c c c } \hline \hline
         & $y<0$ & $0\leq y \leq 1$ & $y>1$ \\ \hline
     $pp$ (exp)& $0.13 \pm 0.05$ & - & $0.18 \pm 0.03$ \\
     $pp$ (QGSM)& ($0.13 \pm 0.04$) & - & ($0.08 \pm 0.01$) \\ \hline
    $\pi^-\pi^-$ (exp)  & $0.26 \pm 0.06$ & $0.21 \pm 0.03$ & $0.13 \pm 0.06$ \\
    $\pi^-\pi^-$ (QGSM)   & ($0.11 \pm 0.01$) & ($0.12 \pm 0.01$) & ($0.08 \pm 0.01$) \\ \hline \hline
 \end{tabular}
  \end{center}
\end{table}
  
  In conclusion, in this paper the two  particle azimuthal correlations
 are studied for $pp$ and $\pi^-\pi^-$ pairs in 4.2 A GeV C+Ta collisions.
 Their dependence on collision centrality, rapidity  and rapidity
 difference is studied. When  both protons in the pair cover the whole
 rapidity range, they show very weak back-to-back correlations,
 while side-by-side correlations are observed  when
 both protons  are restricted to the target or projectile fragmentation region.
 Pions always show side-by-side correlations, with slight decrease in
 magnitude when going from the target to the projectile
 rapidity region. While back-to-back correlations can be attributed to the
 transverse momentum conservation, all side-by-side correlations can be
 attributed to the flow effects since the contribution of quantum correlations
 to the overall azimuthal correlations is negligible. The underlying mechanism
 that  leads to the different magnitudes and dependence on rapidity
 for the directed flow of protons and pions, involves strong rescattering
 and absorption of pions emitted into direction of the huge
 tantalum nucleus. At the later stage of collision, when the
 spectator matter leaves the collision zone, rescattering of
 protons near the beam (target) rapidity region is small,
 while the pions are   influenced by the shadowing effect of the
 participant nucleons trough both pion rescattering and absorption.
 Assuming that  observed side-by-side  correlations can be attributed to
 flow effects, the intensity of the  directed flow is determined from the
 two particle azimuthal correlations  without event-by event estimation of
 the reaction plane. It was found that both two-particle
 azimuthal correlations and Fourier expansion of the azimuthal
 distribution of particles with respect to the reaction plane give
 the consistent prediction for the directed flow of protons and pions.\\

 \end{document}